\documentclass[pra,twocolumn]{revtex4}  
\usepackage{dcolumn}   
\usepackage{bm}        
\usepackage{amssymb}   
\usepackage{textcomp}
\usepackage{amsmath}
\usepackage{color}
\usepackage{braket}

\begin{document}
\title{Optimal quantum preparation contextuality in $n$-bit parity-oblivious multiplexing task}
\author{Shouvik Ghorai$^{1}$ \footnote{ghorai.shouvik@gmail.com}}
\author{A. K. Pan $^{2}$\footnote{akp@nitp.ac.in}}
\affiliation{$^{1}$Indian Institute of Science Education and Research Kolkata, Mohanpur, Nadia 741246, India}
\affiliation{$^{2}$National Institute of Technology Patna, Ashok Rajhpath, Patna 800005, India}

\begin{abstract}
In [ PRL, 102, 010401 (2009)], Spekkens \emph{et al.} have shown that quantum preparation contextuality can power the parity-oblivious multiplexing (POM) task. The bound on the optimal success probability of $n$-bit POM task performed with the classical resources was shown to be the \textit{same} as in a preparation non-contextual theory. This non-contextual bound is violated if the task is performed with quantum resources. While in $2$-bit POM task the optimal quantum success probability is achieved, in $3$-bit case optimality was left as an open question.  In this paper, we show that the quantum success probability of a $n$-bit POM task is solely dependent on a suitable $2^{n-1}\times n$ Bell's inequality and optimal violation of it optimizes the success probability of the said POM task.  Further, we discuss how the degree of quantum preparation contextuality restricts the amount of quantum violations of Bell's inequalities, and consequently the success probability of a POM task.

\end{abstract}
\maketitle

\section{Introduction}
By demonstrating an ingenious \emph{gedanken} experiment  Einstein, Podoloski and Rosen had remarked\cite{epr} that the quantum mechanical description of nature by using $\psi$ function  is inherently incomplete. The ontological models of an operational quantum theory seek to provide a `complete specification' of the state of a quantum system so that the individual measured values of any dynamical variable are predicted by an appropriate set of ontic states (usually denoted as $\lambda$'s). Studies on this issue have resulted in spectacular discoveries about the constraints that need to be imposed on the ontological models in order to be consistent with the statistics of quantum mechanics (QM).  Bell's theorem \cite{bell64} is the first which provides a constraint that an ontological model has to be nonlocal.  Shortly after the Bell's theorem, Kochen and Specker (KS) \cite{bell66,kochen} demonstrated an inconsistency between QM and the non-contextual ontological models.

In a non-contextual ontological model the individual measured values of an observable that occur for an appropriate set of $\lambda$s is irrespective of the way it is being measured. Let an observable $\widehat{A}$ be commuting with $\widehat{B}$ and $\widehat{C}$, with $\widehat{B}$ and $\widehat{C}$  being non-commuting. Then the assumption of non-contextuality asserts that the value occurring in a measurement $\widehat{A}$ is independent of, whether the measurement is performed with $\widehat{B}$ or $\widehat{C}$.  KS theorem \cite{bell66,kochen} demonstrates that such a non-contextual assignment of values is impossible for all possible set of measurements for $d\geq 3$. The original KS proof was demonstrated using 117 projectors for qutrit system. Later, simpler versions of it using lower number of projectors have been provided \cite{ker}. Apart from the KS-type all-versus-nothing proof, inequality based proofs have also been provided \cite{mermin}.  Note that KS-type proof has a limited applicability because it uses an additional assumption of outcome determinism for sharp measurement along with the usual measurement non-contextuality assumption. Also, it is not applicable to any arbitrary operational theory, rather is specific to quantum theory. The traditional notion of non-contextuality was further generalized by Spekkens\cite{spekk05} for any arbitrary operational theory and extended the formulation to the transformation and preparation non-contextuality. The present paper concerns the notion of preparation non-contextuality of an ontological model and its quantum violation. 

Before proceeding further let us recapitulate the essence of an ontological model reproducing the quantum statistics \cite{hari}. Given a preparation procedure $P\in \mathcal{P}$ and a measurement procedures $M\in \mathcal{M}$, an operational theory assigns probability $p(k|P, M)$ of obtaining a particular outcome $k\in \mathcal{K}_{M}$. Here $\mathcal{M}$ is the set of measurement procedures and $\mathcal{P}$ is the set of preparation procedures. In QM, a preparation procedure produces a density matrix $\rho$ and measurement procedure (in general described by a suitable POVM $E_k$) provides the probability of a particular outcome $ k $ is given by $p(k|P, M)=Tr[\rho E_{k}]$, which is the Born rule. In an ontological model of QM, it is assumed that whenever $\rho$ is prepared by a specific preparation procedure $P\in \mathcal{P}$ a probability distribution $\mu_{P}(\lambda|\rho)$ in the ontic space is prepared, satisfying $\int _\Lambda \mu_{P}(\lambda|\rho)d\lambda=1$ where $\lambda \in \Lambda$ and $\Lambda$ is the ontic state space. The probability of obtaining an outcome $k$ is given by a response function $\xi_{M}(k|\lambda, E_{k}) $ satisfying $\sum_{k}\xi_{M}(k|\lambda, E_{k})=1$ where a measurement operator $E_{k}$ is realized through a particular measurement procedure $M\in\mathcal{M}$. A viable ontological model should reproduce the Born rule, i.e., $\forall \rho $, $\forall E_{k}$ and $\forall k$, $\int _\Lambda \mu_{P}(\lambda|\rho) \xi_{M}(k|\lambda, E_{k}) d\lambda =Tr[\rho E_{k}]$.

According to the generalized notion of non-contextuality proposed by  Spekkens \cite{spekk05}, an ontological model of an operational theory can be assumed to be non-contextual if two experimental procedures are operationally equivalent, then they have equivalent representations in the ontological model. If two measurement procedures $ M $ and $ M' $ produces same observable statistics for all possible preparations then the measurements $M$ and $M'$ belong to the equivalent class. An ontological model of QM is assumed to be measurement non-contextual if $\forall P :  p(k|P, M)=p(k|P, M^{\prime})\Rightarrow \xi_{M}(k|\lambda, E_{k})=\xi_{M^{\prime}}(k|\lambda, E_{k})$ is satisfied. KS non-contextuality assumes the aforementioned measurement non-contextuality along with the outcome determinism for the sharp measurement. Similarly, an ontological model of QM can be considered to be preparation non-contextual one if  $\forall M :  p(k|P, M)=p(k|P^{\prime}, M)\Rightarrow \mu_{P}(\lambda|\rho)=\mu_{P^{\prime}}(\lambda|\rho)$ is satisfied where $P$ and $P^{\prime}$ are two distinct preparation procedures but in the same equivalent class. In an ontological model of QM, the preparation non-contextuality implies the outcome determinism for sharp measurements and  also implies KS non-contextuality \cite{spekk05}. Then any KS proof can be considered as a proof of preparation contextuality but converse does not hold. In this sense, preparation non-contextuality is a stronger notion than traditional KS non-contextuality \cite{leifer}. Very recently, it is also shown \cite{kunj16} that any ontological model satisfying both the assumptions of preparation and the measurement non-contextuality  cannot reproduce all quantum statistics, even if the assumption of outcome determinism for sharp measurement is dropped. Experimental test of such an universal non-contextuality has also been provided which are free from idealized assumptions of  noiseless measurements and exact operational equivalences \cite{mazurek}.   

The failure of non-contextuality is a signature of non-classicality which is of foundational importance. It would then be interesting if this non-classical feature of contextuality can be used as a resource for providing advantage in various information processing and computation tasks, similar to the spirit of the violations of Bell inequalities which have been shown to be a resource for device-independent key distribution \cite{barrett} and certified randomness \cite{acin16}.  In an interesting work Spekkens \emph{et al.} \cite{spekk09} have first demonstrated how quantum preparation contextuality can power a communication game termed as parity-oblivious multiplexing (POM) task. 

The essence of a $n$-bit POM task can be encapsulated as follows. Alice has a $ n $-bit string $ x $ chosen uniformly at random from $\{0,1\}^n$. Bob can choose any bit $ y \in \{1,2, ..., n\}$ and recover the bit $x_y$ with a probability. The condition of the task is, Bob's output must be the bit $b=x_y$, i.e., the $ y^{th} $ bit of Alice's input string $x$. In other words, Alice and Bob try to optimize the probability $p(b=x_y)$ with the constraint that  \emph{no} information about any parity of $ x $ can be transmitted to Bob. 

\color{black}In \cite{spekk09}, it is shown that a $n$-bit POM task performed with classical resources is constrained by an inequality. Interestingly, the same inequality can be obtained in any ontological model satisfying preparation non-contextuality. For the case of 2-bit POM task they have achieved the optimal quantum success probability, which is recently reaffirmed \cite{banik14} through the Cirelson bound. For 3-bit POM task, Spekkens \emph{et al.} \cite{spekk09} provided the quantum advantage over the non-contextual POM task but the question pertaining to its optimality is left as an open problem. In this paper, we first show that the quantum success probability provided in \cite{spekk09} is indeed the optimal one for the $3-$bit POM task. In order to demonstrate this we first prove that the optimal success probability of $3-$bit POM task in QM is solely dependent on the optimal quantum violation of the \textit{elegant Bell inequality} proposed by Gisin \cite{gisin}. We then generalize our approach to show that one can find a suitable $2^{n-1}\times n$ elegant Bell's inequality for $n$-bit POM task  and optimal violation of it provides the optimal success probability of the POM task. Further, we discuss how the degree of quantum preparation contextuality puts constraint on the quantum violations of Bell's inequalities and consequently on the quantum success probability of a POM task. 

\section{The POM task and preparation non-contextuality}

Following \cite{spekk09}, we define a parity set $ \mathbb{P}_n= \{x|x \in \{0,1\}^n,\sum_{r} x_{r} \geq 2\} $ with $r\in \{1,2,...,n\}$. The cryptographic constraint here is the following. For any $s \in \mathbb{P}_{n}$, no information about $s.x = \oplus_{r} s_{r}x_{r}$ (s-parity) is to be transmitted to Bob, where $\oplus$ is sum modulo $ 2 $. For example, when $ n=2 $ the set is $\mathbb{P}=\{11\} $, so no information about $x_1\oplus x_2$ can be transmitted by Alice.

The maximum probability of success in a classical $ n $-bit POM task is ${(n+1)}/{2n}$, because only those encoding of $x$ which does not provide any information about parity are those which encodes a single bit. While the explicit proof can be found in \cite{spekk09}, a simple trick can saturate the bound as follows. Assume that Alice always encodes the first bit (pre-discussed between Alice and Bob) and sends to Bob.  If $y=1$, occurring with probability $ 1/n $, Bob can predict the outcome with certainty and for $y \neq 1$, occurring with probability of $(n-1)/n$, he at best guesses the bit with probability $1/2$. Hence the total probability of success is ${1}/{n} + {(n-1)}/{2n} = {(n+1)}/{2n}$. Since $y$ is chosen uniformly it is irrelevant which bit is encoded. This does not disclose the parity information to Bob. 

Let in an operational theory, Alice encodes her $ n $-bit string of $ x $ prepared by a procedure $P_{x}$. Next, after receiving the particle , for every $y \in \{1,2,...,n\}$, Bob performs a two-outcome measurement $M_{y}$ and reports outcome $b$ as his output. Then the probability of success is given by
\begin{equation}
\label{qprob}
	p(b=x_{y}) = \dfrac{1}{2^n n}\sum\limits_{x,y}^{}p(b=x_y|P_{x},M_{y}).
\end{equation}

The parity-obliviousness condition in the operational theory guarantees that there is no outcome of any measurement  for which the probabilities for s-parity $0$ and s-parity $1$ are different. Mathematically,
\begin{equation}
\label{poc}
	\forall s \; \forall M \; \forall k \sum\limits_{x|x.s=0} p(P_x|k,M) = \sum\limits_{x|x.s=1} p(P_x|k,M).
\end{equation} 

For a preparation non-contextual ontological model, it is proved \cite{spekk09} that the success probability in $n$-bit POM  task satisfies the bound
\begin{equation}
	p(b = x_y)_{pnc} \leq \dfrac{1}{2}\left(1+\dfrac{1}{n}\right).
\end{equation}

In order to derive this bound it is proved that in a preparation non-contextual models, parity-obliviousness at the operational level implies equivalent representation in the ontological model, so that, $\forall M: \sum\limits_{x|x.s=0} \mu(\lambda|P_{x},M)=\sum\limits_{x|x.s=1} \mu(\lambda|P_{x},M)$ is satisfied. This is true even if Bob can perfectly determine the ontic state $\lambda $.

In quantum POM task, Alice encodes her $ n $-bit string of $ x $ into pure quantum states $ \rho_{x} =|\psi_{x}\rangle\langle\psi_{x}|$, prepared by a procedure $P_{x}$. After receiving the particle, Bob performs a two-outcome measurement $M_{y}$ for every $y \in \{1,2,...,n\}$ and reports outcome $b$ as his output. Spekkens \emph{et al.} \cite{spekk09} have proved that the optimal quantum success probability for $2$-bit POM task is $p^{opt}_{Q}=(1/2)(1+1/\sqrt{2}) >p(b = x_y)_{pnc}=3/4$ . For $3-$bit quantum POM task they provided a success probability $p_{Q}=(1/2)(1+1/\sqrt{3}) $  but left open the question of optimality of it.  

Recently,  Chailloux \emph{et al.} \cite{chali} have shown that for even POM task optimal success probability  is $(1/2)(1+1/\sqrt{n})$. However, they first proved that POM task can be shown to be equivalent to an another game in some conditions and then optimize the success probability of that game. By taking a different approach, Banik \emph{et al.}\cite{banik15} obtained the quantum optimal success probability of the $2-$bit POM task through the Tsirelson bound \cite{cirel} of CHSH inequality \cite{chsh}. Instead of two-outcome measurement, Hameedi \emph{et al.} \cite{hameedi} have derived the non-contextual bound for $m$-outcome scenario is given by  $p(b = x_y)_{pnc}\leq (n + m-1)/(n m)$. However, they have numerically optimized the quantum success probability of POM task for $n=2$ and $m=3...7$.   

We use the similar direct approach adopted in \cite{banik15} to derive the optimal success probability of $n-$bit POM task for dichotomic outcomes. Interestingly, the success probability can be  shown to be solely linked to the $2^{n-1}\times n$ elegant Bell's inequality \cite{gisin} which obviously reduces to  the CHSH inequality  for $2$-bit POM task. Further, by using an interesting technique we analytically optimize $2^{n-1}\times n$ elegant Bell's inequality which in turn provides the optimal quantum success probability ($p^{opt}_{Q}$) of $n$-bit POM task. In order to showing this, let us first provide an explicit derivation of $p^{opt}_{Q}$ for 3-bit POM task which will help the reader to understand the optimization of $p_{Q}$  for $n$-bit quantum POM task.

\section{ 3-bit POM task and optimal quantum success probability}

For 3-bit POM task Alice chooses her bit $x$ randomly from $\{0,1\}^{3}$. We rewrite all the possible $x$ as an ordered set 
$\mathcal{D}_3 = (000, 001, 010, 100, 011, 101, 110, 111)$.

The parity set is then $ \mathbb{P}_3 = \{011, 101, 110, 111\} $.  If we consider the case when $ s=110 $, the bits having $s-$parity 0 are \{000, 001, 110, 111\} and  the bits have $s-$parity 1 are \{010, 100, 011, 101\}. From Eq. \eqref{poc}, the parity-obliviousness in a non-contextual ontological model can then be ensured if $  \forall M $ and $ \forall k $
\begin{align}
\label{po}
p(&P_{000}|k,M) + p(P_{001}|k,M) + p(P_{110}|k,M)\nonumber \\ &  +p (P_{111}|k,M)   
= 	p(P_{010}|k,M) + p(P_{100}|k,M)\nonumber \\ & + p(P_{011}|k,M) +p (P_{101}|k,M)  .
\end{align} 

Similar parity-oblivious conditions can be found for every other element of $ \mathbb{P}_{3}$. 

Let us consider an entangled quantum state $\rho_{AB}=|\psi_{AB}\rangle\langle\psi_{AB}|$ where $|\psi_{AB}\rangle \in \mathcal{C}^{2}\otimes\mathcal{C}^{2}$. Alice randomly performs one of the four projective measurements $\{P_{A_{i}},I-P_{A_{i}}\}$ where $i=1,2,3,4$ to encode her input $x$ into eight pure qubits as $ \rho_{x}$, are given by 
\begin{subequations}	
\begin{align}
	\dfrac{1}{2} \rho_{000} & = Tr_1\Big[(P_{A_1} \otimes I) \; \rho_{AB}\Big]\\
	\dfrac{1}{2} \rho_{111} & = Tr_1\Big[(I-P_{A_1} \otimes I) \; \rho_{AB}\Big]\\
	\dfrac{1}{2} \rho_{001} & = Tr_1\Big[(P_{A_2} \otimes I) \; \rho_{AB}\Big]\\
	\dfrac{1}{2} \rho_{110} & = Tr_1\Big[(I-P_{A_2}\otimes I) \; \rho_{AB}\Big]\\
	\dfrac{1}{2} \rho_{010} & = Tr_1\Big[(P_{A_3} \otimes I) \; \rho_{AB}\Big]\\
	\dfrac{1}{2} \rho_{101} & = Tr_1\Big[(I-P_{A_3} \otimes I) \; \rho_{AB}\Big]\\ 
	\dfrac{1}{2} \rho_{100} & = Tr_1\Big[(P_{A_4} \otimes I) \; \rho_{AB}\Big]\\
	\dfrac{1}{2} \rho_{011} & = Tr_1\Big[(I-P_{A_4} \otimes I) \; \rho_{AB}\Big].
\end{align}
\end{subequations}

After receiving the information from Alice, Bob performs three projective measurements $\{P_{B_{y}},I-P_{B_{y}}\}$ with  $y=1,2,3$.

From the construction we have, $ \frac{1}{2}\rho_{000} + \frac{1}{2}\rho_{111} = \frac{1}{2}\rho_{110} + \frac{1}{2}\rho_{001} =  \frac{1}{2}\rho_{010} + \frac{1}{2}\rho_{101} = \frac{1}{2}\rho_{011} + \frac{1}{2}\rho_{100} = {\mathbb{I}}/{2}$. So, the parity oblivious constraint in QM is satisfied if $\frac{1}{4}(\rho_{000} + \rho_{111} + \rho_{110} + \rho_{001}) =\frac{1}{4}( \rho_{010} + \rho_{101} + \rho_{011} + \rho_{100})$.

Spekkens \emph{et al.}\cite{spekk09} has obtained a quantum success probability $ p_{Q} = (1/2)\big(1+1/\sqrt{3}\big) $ of the 3-bit POM task. Given an entangled state $ \ket{\Psi_{AB}} = (\ket{00} + \ket{11})/\sqrt{2}$, if the following choices of observables in Alice's end are made, so that, $A_1 = (\sigma_x + \sigma_y + \sigma_z )/\sqrt{3}$, $A_2 = (\sigma_x + \sigma_y - \sigma_z )/\sqrt{3}$, $A_3 = (\sigma_x - \sigma_y + \sigma_z) /\sqrt{3}$ and $A_4 = (-\sigma_x + \sigma_y + \sigma_z )/\sqrt{3}$, and Bob chooses $B_1 = \sigma_x$,  $B_2 = -\sigma_y$ and $B_3 = \sigma_z$, then the above bound can be achieved.  Similarly, one may chose another entangled state for which a different set of observables is required to obtain that bound. The question is whether the above quantum success probability is optimal. 

In this paper, we first prove that quantum success probability for $3-$bit POM task obtained by Spekkens \emph{et al.}\cite{spekk09} is indeed the optimal one. This is shown through the optimal violation of the elegant Bell inequality \cite{gisin}. In order to showing this, let us explicitly write down the quantum success probability for $3$-bit POM task by using Eq.(\ref{qprob}) is given by
\begin{eqnarray}
\label{prob}
&&p_{Q} =  \dfrac{1}{24}  \Big[ p(0|\rho_{000},P_{B_1})+p(0|\rho_{000},P_{B_2})+p(0|\rho_{000},P_{B_3}) \nonumber \\ &&+p(0|\rho_{001},P_{B_1})+p(0|\rho_{001},P_{B_2})+p(0|\rho_{001},I-P_{B_3})  \nonumber  \\  &&+p(0|\rho_{010},P_{B_1})+p(1|\rho_{010},I-P_{B_2})+p(0|\rho_{010},P_{B_3})  \nonumber  \\  &&+p(1|\rho_{100},I-P_{B_1})+p(0|\rho_{100},P_{B_2})+p(0|\rho_{100},P_{B_3})  \nonumber  \\  &&+p(0|\rho_{011},P_{B_1})+p(1|\rho_{011},I-P_{B_2})+p(1|\rho_{011},I-P_{B_3})\nonumber  \\  &&+p(1|\rho_{101},I-P_{B_1})+p(0|\rho_{101},P_{B_2})+p(1|\rho_{101},I-P_{B_3})\nonumber  \\  &&+p(1|\rho_{110},I-P_{B_1})+p(1|\rho_{110},I-P_{B_2})+p(0|\rho_{110},P_{B_3})\nonumber  \\  && + p(1|\rho_{111},I-P_{B_1})+p(1|\rho_{111},I-P_{B_2}) \nonumber  \\ &&+p(1|\rho_{111},I-P_{B_3}) \Big].
\end{eqnarray}

Further simplification and rearrangements provide the following form is given by
\begin{align}
\label{prob1}
p_{Q} = \dfrac{1}{2} + \dfrac{\langle\mathcal{B}_3\rangle}{24} 
\end{align}
where $\mathcal{B}_3$ is the elegant Bell expression\cite{gisin} is given by
\begin{align}
\label{elegant}
\mathcal{B}_3 = &(A_1 + A_2 + A_3 - A_4)\otimes B_1 \nonumber \\
+ & (A_1 + A_2 - A_3 + A_4)\otimes B_2 
\\ + & (A_1 - A_2 + A_3 + A_4)\otimes B_3
\nonumber
\end{align} 
The detailed calculation to derive Eq.(\ref{prob1}) from Eq.(\ref{prob}) is shown in the Appendix A. 

We have thus shown that  the optimality of $p_{Q}$ for 3-bit POM task requires the optimal violation of elegant Bell inequality. For this, by following \cite{acin}, we define $ \gamma_3 = 4\sqrt{3}\, \mathbb{I} - \mathcal{B}_3 $. Since $A_i^\dagger A_i = \mathbb{I} = B_y^\dagger B_y$, $ \gamma_3 $ can be decomposed as $ \gamma_3 = (\sqrt{3}/2) \sum_{i=1}^{4} M_i^\dagger M_i $ where $ M_i $'s are linear combination of $ A_i $'s and $ B_y $'s
\begin{align}
	M_1 & = (B_1+B_2+B_3)/\sqrt{3} -A_1 \nonumber \\
	M_2 & = (B_1+B_2-B_3)/\sqrt{3} -A_2 \nonumber \\
	M_3 & = (B_1-B_2+B_3)/\sqrt{3} -A_3 \nonumber \\
	M_4 & = (-B_1+B_2+B_3)/\sqrt{3} -A_4 .
\end{align}

Since $\gamma_3$ is positive semi-definite, we have  $ \langle\mathcal{B}_3\rangle^{opt} = 4\sqrt{3}$. This in turn optimize the success probability given by Eq.(\ref{prob1}), so that, $p^{opt}_{Q} = (1/2)(1+(1/\sqrt{3}))$ for a $3-$bit POM task. Thus, $p^{opt}_{Q}$ of $3-$bit POM task is achieved through the optimal quantum violation of the elegant Bell's inequality. 

It is interesting to note here that $\langle\mathcal{B}_3\rangle$ can be saturated if the choice of $B_{y}$'s can be made in the following way, so that  $A_1 + A_2 + A_3 - A_4=(4/\sqrt{3}) B_{1}$, $A_1 + A_2 - A_3 + A_4=(4/\sqrt{3}) B_{2}$ and $A_1 - A_2 + A_3 +A_4=(4/\sqrt{3}) B_{3}$. Then, $\langle\mathcal{B}_3\rangle=(4/\sqrt{3})\sum_{y=1}^{3} \langle B_y\otimes B_y\rangle$ provides $\langle\mathcal{B}_3\rangle^{opt}=4\sqrt{3}$ provided for a suitable state, each of the $\langle B_y\otimes B_y\rangle$ is equal to 1. The important question is whether such a choice of the observables and the state can be found. In fact, the choice made by in \cite{spekk09} satisfies the above requirements.   

Note that, the algebraic maximum of Eq.(\ref{elegant}) is 12 which may be obtained for a post-quantum theory (PR box is an example for the case of Bell-CHSH expression) providing the maximum violation of parity-obliviousness condition. 

The above calculation is performed by assuming element $s=110$ from the set $\mathbb{P}_{3}$. One may take any of the other three elements of $\mathbb{P}_3$ to find the optimal success probability. However,  $p^{opt}_{Q}$ will remain same for any of such cases.  We now proceed to generalize the approach for $n$-bit POM task.


\section{$n$-bit POM task and optimal success probability in QM}
For $n$-bit POM task Alice chooses her bit $x^{\delta}$ randomly from $\{0,1\}^{n}$ with $\delta\in \{1,2...2^{n}\}$. The relevant ordered set $\mathcal{D}_n$ can be written as 
$\mathcal{D}_n = (x^{\delta} | x^i \oplus x^j = 111...11 \;\text{and}\; i+j = 2^n +1 ) $
and $i \in \{1,2, ...2^{n-1}\}$. Here, $ x^1 = 00...00, x^2 = 00...01, .... $, and so on. The parity set is defined as
$\mathcal{P}_{n}= \{x^{\delta}|x^{\delta} \in \{0,1\}^n,\sum_{r} x^{\delta}_{r} \geq 2\} $. We choose $x^s = 1100...00$ and fix the \textit{s-parity 0} and \textit{s-parity 1} sets. 

Let us consider a suitable entangled state $ \rho_{AB} = \ket{\psi_{AB}}\bra{\psi_{AB}}$ with $|\psi_{AB}\rangle \in \mathcal{C}^{d}\otimes\mathcal{C}^{d}$. Alice performs one of the $2^{n-1}$ projective measurements $\{{P_{A_i}}, \mathbb{I}-P_{A_i}\}$ where $ i \in \{1, 2, ... 2^{n-1}\}$ to encode her $n$-bits into $ 2^n $ pure quantum states are given by
\begin{subequations}
	\begin{align}
	\dfrac{1}{2} \rho_{x^i} &= tr_A[(P_{A_i} \otimes \mathbb{I}) \rho_{AB}]\\
	\dfrac{1}{2} \rho_{x^j} &= tr_A[(\mathbb{I}-P_{A_i} \otimes \mathbb{I}) \rho_{AB}]; 
	\end{align}
\end{subequations}
with $ i+j=2^n+1 $.

We define Bob's measurements as 
\begin{align}
	M_y = \begin{cases}
	M_y^i,\text{when}\; b=x_y^i \\ M_y^j, \text{when}\; b=x_y^j
	\end{cases}\\
	M_{y}^{i(j)} = \begin{cases}
	P_{B_y},& \text{when}\; x^{i(j)}_y=0 \ \\ \mathbb{I} - P_{B_y},& \text{when}\; x^{i(j)}_y=1 
	\end{cases}	
\end{align}
The quantum success probability can then be written as 
\begin{align}
	p_Q &=  \dfrac{1}{2^n n} \sum_{y=1}^{n} \sum\limits_{i=1}^{2^{n-1}} p(b=x^i_y|\rho_{x^i},M_{y}^i) \nonumber \\ & \quad + p(b=x^j_y|\rho_{x^j},M_{y}^j)
\nonumber \\ & 
\label{qprobn}
=\dfrac{1}{2^n n} \sum_{y=1}^{n} \sum\limits_{i=1}^{2^{n-1}} tr[\rho_{x^i} M_{y}^i] + tr[\rho_{x^j} M_{y}^j]
\end{align} 

Since  $ \forall i,j $, $x^i \oplus x^j = 111...111$ we have $x^i_y \oplus x^j_y = 1$. Then, while $ x^i_y=0 $ we can write $ tr[\rho_{x^i} M_{y}^{i}] + tr[\rho_{x^j} M_{y}^{j}] =tr[\rho_{x^j}] + tr[(\rho_{x^i} - \rho_{x^j})P_{B_y}]$, and while $ x^j_y=0 $ we have $tr[\rho_{x^i} M_{y}^{i}] + tr[\rho_{x^j} M_{y}^{j}] = tr[\rho_{x^i}] - tr[(\rho_{x^i} - \rho_{x^j})P_{B_y}]$. Hence, if $ x^i_y=0$ the term $tr[\rho_{x^j}] $ exists, while $x^i_y=1,$ then $tr[\rho_{x^i}] $ exists. So, Eq.\eqref{qprobn} can be written as

\begin{align}
\label{gisinn}
p_Q &= \dfrac{1}{2^n n} \sum_{y=1}^{n} \sum\limits_{i=1}^{2^{n-1}} (-1)^{x^i_y} tr[(\rho_{x^i} -\rho_{x^j}) P_{B_y}] + tr[\rho_{x^{(i. x_{y}^{i}+j.x_{y}^{j})}}] \nonumber \\
&=\dfrac{2^{n-1}n}{2^n n} + \dfrac{1}{2^n n} \sum_{y=1}^{n}\sum_{i=1}^{2^{n-1}} (-1)^{x_y^i} \langle (2 P_{A_i}- I) \otimes 2 P_{B_y} \rangle \nonumber \\
&= \dfrac{1}{2} + \dfrac{1}{2^n n} \sum_{y=1}^{n}\sum_{i=1}^{2^{n-1}} (-1)^{x^i_y}  \langle A_i\otimes B_y\rangle 
\end{align}

Then the  success probability of $n$-bit POM task is dependent on the $2^{n-1}\times n$ elegant Bell expression 

\begin{align}
	\mathcal{B}_{n} =  \sum_{y=1}^{n}\sum_{i=1}^{2^{n-1}} (-1)^{x^i_y}  A_i\otimes B_y
\end{align}

In order to optimize $ \mathcal{B}_{n} $ we define $\gamma_n = 2^{n-1}\sqrt{n}\, I - \mathcal{B}_{n}$. By considering $ A_i^\dagger A_i = I = B_y^\dagger B_y $, $ \gamma_{n} $ can be written in the following way $\gamma_{n} = \dfrac{\sqrt{n}}{2} \sum_{i=1}^{2^{n-1}} M_i^{\dagger} M_i$ where $M_i = \sum_{y}(-1)^{x^{i}_y} \dfrac{B_y}{\sqrt{n}} - A_i $. Since $\gamma_{n}\geq 0$ we have

\begin{align}
	\mathcal{B}_{n} \leq 2^{n-1}\sqrt{n}\, \mathbb{I}
\end{align}

It is then straightforward to see from Eq.(\ref{gisinn}) that the optimal quantum success probability for $n-$bit POM task is 
\begin{equation}
\label{nopt}
	p_Q^{opt} = \dfrac{1}{2} \Big(1+\dfrac{1}{\sqrt{n}} \Big).
\end{equation}

Thus, $p_Q^{opt} \geq p_{pnc}^{opt}$ for $n$-bit POM task. The question remains whether such an amount of success probability can be achieved for any $n$ if Alice uses qubit system for encoding her input into pure states. Clearly, if the choices of observables is found for which $(2^{n-1}/\sqrt{n})\sum_{i=1}^{2^{n-1}} (-1)^{x^i_y}  A_i= B_y$ is satisfied then we have $\langle\mathcal{B}_{n}\rangle =  \sum_{y=1}^{n}\langle B_y\otimes B_y\rangle$ which may provide $\mathcal{B}_{n}^{opt}=2^{n-1}\sqrt{n}$ provided each of the $\langle B_y\otimes B_y\rangle=1$. We have already shown that for $n=2$ and $3$ such choices of observables are available for qubit system. However, for $n>3$ the observables cannot be found in qubit system to obtain the optimal quantum bound. We provide an explicit example in Appendix B to show that how optimal quantum success probability of 4-bit POM task can be achieved when Alice uses two-qubit system for encoding her input. In Appendix C, by following \cite{chali}, we write down the explicit construction of such set of observables for which the violation is optimal. However, the dimension of the Hilbert space needs to $[2^{n/2}]$ for $n-$bit POM task.
\section{Summary and Discussions}

We studied how the quantum preparation contextuality provides advantage in a POM task. The success probability of the $n$-bit POM task is shown to exceed the non-contextual bound if performed with quantum resources. Spekkens \emph{et al.} \cite{spekk09} have provided the optimal quantum success probability of $2$-bit POM task which is reaffirmed \cite{banik15} through the Cirelson bound of CHSH inequality.  The $p_{Q}$ of $3$-bit POM task is shown \cite{spekk09} to be larger than non-contextual bound but optimality  of it was left as an open question. 

By using an interesting approach, we showed that the success probability of a $n$-bit POM task can be solely dependent on the quantum violation of  $2^{n-1}\times n$ Bell's inequality. Thus, the derivation of $p_Q^{opt}$ of $n-$bit POM task reduces to the optimization of the relevant Bell expression. For $n=2$, the Bell inequality is the CHSH one and for $n=3$ we have the elegant Bell's inequality \cite{gisin}. By using an interesting technique \cite{acin}, we first optimize the elegant Bell expression arising from $3$-bit POM task and further generalized it for $n$-bit case. The optimal quantum value of  $2^{n-1}\times n$ elegant Bell's expression is $2^{n-1}\sqrt{n}$ which in turn provides the optimal success probability $p^{opt}_{Q}=(1/2)(1+1/\sqrt{n}) $ for $n$-bit POM task. Note that for $n=2,3$, $p^{opt}_{Q} $ can be obtained even if Alice chooses pure qubit states for encoding her bits. But, for $n>3$ the encoding by using pure qubits does not provide the optimal success probability in QM. In Appendix B, we showed that for $4$-bit task Alice's encoding in two-qubit pure state succeeds in achieving the optimal success probability. We provide the general construction of observables and the required dimension of entangled state in order to obtain $p^{opt}_{Q} $ for $n$-bit POM task in Appendix C.  


Note that, the success probability can be unity if the value of relevant Bell expression reaches to its algebraic maximum. However, such amount of violation of Bell's inequality may be  obtained in a post-quantum theory which then implies the highest degree of preparation contextuality. In such a case, the overlap between the respective probability distributions $\mu(\lambda|\rho_{x|x.s=0})$ and $\mu(\lambda|\rho_{x|x.s=1})$ corresponding to s-parity 0 and s-parity 1 requires to be maximum. In QM, the maximum success probability is $(1/2)(1+1/\sqrt{2}) $, that is for $2$-bit POM task. Then the highest degree of preparation contextuality is not allowed in QM. Note here that although the success probability $p_{Q}^{opt}\geq p(b = x_i)_{pnc}$ for any $n>2$ but $p_{Q}^{opt}$ decreases with the increment of the number of bit $n$. The effect of preparation contextuality is then prominent here.  The condition of parity-obliviousness produce two mixed states in Bob's side and such preparation procedures fix the relevant Bell's inequality. Then, the overlap between  $\mu(\lambda|\rho_{x|x.s=0})$ and $\mu(\lambda|\rho_{x|x.s=1})$ in the ontic space $\Lambda$ for the case of $2-$bit POM task is larger than $3-$bit case. It is then straightforward to understand that for $n-$bit POM tasks,  both $\mu(\lambda|\rho_{x|x.s=0})$ and $\mu(\lambda|\rho_{x|x.s=1})$ contains  distributions corresponding to $2^{n-1}$ pure states, so that, every pure state in $\rho_{x|x.s=0}$ is very much close to a pure state in $\rho_{x|x.s=1}$ yielding the distributions of ontic states for the mixed state indistinguishable (i.e., preparation noncontextual) in the ontic space which thereby providing the lowest  success probability. Thus, optimal quantum preparation contextuality limits the amount of violation of Bell's inequality and fixes the maximum success probability of the POM task.

\section*{Acknowledgments}

Authors thanks G. Kar for insightful discussions. SG acknowledges the local hospitality of NIT Patna during his visit. AKP acknowledges the support from Ramanujan Fellowship Research Grant (SB/S2/RJN-083/2014).

\begin{widetext}
\appendix

\section{}

Explicit derivation of Eq.(\ref{prob1}) in the main text is shown. The quantum success probability given by Eq.(\ref{prob}) can be rearranged as 
\begin{eqnarray}
\label{a2}
p_{Q} &=& \dfrac{1}{24}\Big( Tr[\rho_{001}] + Tr[\rho_{010}] + Tr[\rho_{100}] + 2\;Tr[\rho_{011}] + 2\;Tr[\rho_{101}] + 2\;Tr[\rho_{110}] + 3\;Tr[\rho_{111}]\nonumber\\
&+& Tr\big[(\rho_{000}-\rho_{111})P_{B_1} \big] + Tr\big[(\rho_{001}-\rho_{110})P_{B_1} \big] + Tr\big[(\rho_{010}-\rho_{101})P_{B_1} \big] - Tr\big[(\rho_{100}-\rho_{011})P_{B_1} \big]\nonumber\\
&+& Tr\big[(\rho_{000}-\rho_{111})P_{B_2} \big] + Tr\big[(\rho_{001}-\rho_{110})P_{B_2} \big] - Tr\big[(\rho_{101}-\rho_{010})P_{B_2} \big] + Tr\big[(\rho_{100}-\rho_{011})P_{B_2} \big]\nonumber \\
&+& Tr\big[(\rho_{000}-\rho_{111})P_{B_3} \big] - Tr\big[(\rho_{001}-\rho_{110})P_{B_3} \big] + Tr\big[(\rho_{010}-\rho_{101})P_{B_3} \big] + Tr\big[(\rho_{100}-\rho_{011})P_{B_3} \big] \Big)
\end{eqnarray}

Since $\rho_{000}=2 Tr_1\Big[(P_{A_1} \otimes I) \; \rho_{AB}\Big] $ and $\rho_{111}=2 Tr_1\Big[(I- P_{A_1} \otimes I) \; \rho_{AB}\Big] $, we can write $Tr[(\rho_{000}-\rho_{111})P_{B_{1}}]=2 Tr \Big[(2 P_{A_1}-I) \otimes P_{B_{1}}) \Big] $, $Tr[(\rho_{000}+\rho_{111})]=2 Tr \Big[ I \otimes I \Big] = 2 $ and $Tr[(\rho_{000}-\rho_{111})]=2 Tr \Big[(2 P_{A_1}-I) \otimes I \Big]. $

Similarly, writing other terms in Eq.(\ref{a2}), we get the following expression of the success probability in QM is given by
\begin{align}
p_{Q} = \dfrac{1}{24} \Big( Tr[\rho_{100}+\rho_{011}]+Tr[\rho_{010}+\rho_{101}]+Tr[\rho_{001}+\rho_{110}]+ Tr[\rho_{011}] + Tr[\rho_{101}] + Tr[\rho_{110}] + 3\;Tr[\rho_{111}]\Big) \nonumber \\
+ \dfrac{1}{12} \Big[ \big<2P_{A_1}-I \otimes P_{B_1}\big> + \big<2P_{A_2}-I \otimes P_{B_1}\big>  + \big<2P_{A_3}-I \otimes P_{B_1}\big> - \big<2P_{A_4}-I \otimes P_{B_1} \big>  \nonumber \\
+\big<2P_{A_1}-I \otimes P_{B_2}\big> + \big<2P_{A_2}-I \otimes P_{B_2}\big>  - \big<2P_{A_3}-I \otimes P_{B_2}\big> + \big<2P_{A_4}-I \otimes P_{B_2} \big>  \nonumber \\
+\big<2P_{A_1}-I \otimes P_{B_3}\big> - \big<2P_{A_2}-I \otimes P_{B_3}\big> + \big<2P_{A_3}-I \otimes P_{B_3}\big> + \big<2P_{A_4}-I \otimes P_{B_3} \big> \Big]
\end{align}

So, the success probability can be written as
\begin{eqnarray}
\label{last}
p_{Q} &=&\dfrac{1}{2}+ \dfrac{1}{24} \Big[ \big< A_1 \otimes B_1\big> + \big< A_2 \otimes B_1\big> + \big< A_3 \otimes B_1\big> 	- \big<A_4 \otimes B_1 \big> 
+ \big< A_1 \otimes B_2 \big> + \big< A_2 \otimes B_2\big> \nonumber \\
&-& \big< A_3 \otimes B_2 \big> + \big< A_4 \otimes B_2 \big> 
+ \big< A_1 \otimes B_3 \big> - \big< A_2 \otimes B_3 \big> + \big< A_3 \otimes B_3 \big> + \big< A_4 \otimes B_3 \big> \Big]\\
\label{a4}
&=&\dfrac{1}{2} + \dfrac{\langle\mathcal{B}_3\rangle}{24} 
\end{eqnarray}
where  $B_{1}=2 P_{B_{1}}- I$ and  $A_{1}=2 P_{A_{1}}- I$ are used. Eq.(\ref{a4}) is the Eq.(\ref{prob1}) in the main text.

\section{}
In this section, we provide explicit derivation of the optimal quantum success probability of $4$-bit POM task. Similar to $3-$bit case, let us define an ordered set $ \mathcal{D}_4 $, where all possible $ x $'s are written as $\mathcal{D}_4 = (0000, 0001, 0010, 0100, 1000, 0011, 0101, 0110, 1001, 1010, 1100, 0111, 1011, 1101, 1110, 1111)$. The set can be rewritten as $\mathcal{D}_4 = (x^1, x^2, ... x^{16}| x^i + x^j = 1111 \;\text{and}\; i+j = 17) $. We have the following parity set $\mathbb{P}_4 = \{0011,0101,0110,1100,0111,1011,1101,1110,1111\}$ and for our purpose we take $s = 1100$. For $s$-parity $0$ set, Alice encodes her inputs in the following pure states are given by

\begin{subequations}	
	\begin{align}
	\dfrac{1}{2} \rho_{0000} & = Tr_1\Big[(P_{A_1} \otimes I)  \rho_{12}\Big]; \ \ \ 	\dfrac{1}{2} \rho_{1111}  = Tr_1\Big[(I-P_{A_1} \otimes I)  \rho_{12}\Big]\\
	\dfrac{1}{2} \rho_{0001} & = Tr_1\Big[(P_{A_2} \otimes I) \rho_{12}\Big];	\ \ \  \dfrac{1}{2} \rho_{1110}  = Tr_1\Big[(I-P_{A_2}\otimes I) \rho_{12}\Big]\\
	\dfrac{1}{2} \rho_{0010} & = Tr_1\Big[(P_{A_3} \otimes I) \rho_{12}\Big];	\ \ \  \dfrac{1}{2} \rho_{1101}  = Tr_1\Big[(I-P_{A_3} \otimes I)  \rho_{12}\Big]\\ 
	\dfrac{1}{2} \rho_{0011} & = Tr_1\Big[(P_{A_6} \otimes I)  \rho_{12}\Big];	\ \ \  \dfrac{1}{2} \rho_{1100}  = Tr_1\Big[(I-P_{A_6} \otimes I)  \rho_{12}\Big].
	\end{align}
\end{subequations}
and similar encoding for the $s$-parity 1 set. The quantum success probability can be calculated as 

\begin{align}
	p_{Q} = \dfrac{1}{2} + \dfrac{\langle\mathcal{B}_4\rangle}{64} 
\end{align}
where 
$\mathcal{B}_4=(A_1 + A_2 + A_3 + A_4 - A_5 + A_6 + A_7 + A_8)\otimes B_1 +(A_1 + A_2 + A_3 - A_4 + A_5 + A_6 - A_7 - A_8)\otimes B_2+(A_1 + A_2 - A_3 + A_4 + A_5 - A_6 + A_7 - A_8)\otimes B_3+(A_1 - A_2 + A_3 + A_4 + A_5 - A_6 - A_7 + A_8)\otimes B_4$
If we define, 
\begin{align}
\dfrac{1}{4}(A_1 + A_2 + A_3 + A_4 - A_5 + A_6 + A_7 + A_8)\otimes \mathbb{I} =	B_1\otimes \mathbb{I} \nonumber \\
	\label{b3}
	 \dfrac{1}{4}(A_1 + A_2 + A_3 - A_4 + A_5 + A_6 - A_7 - A_8) \otimes \mathbb{I}=B_2\otimes \mathbb{I}  \\
	\dfrac{1}{4}(A_1 + A_2 - A_3 + A_4 + A_5 - A_6 + A_7 - A_8)\otimes \mathbb{I}=B_3\otimes \mathbb{I} \nonumber \\
	\dfrac{1}{4}(A_1 - A_2 + A_3 + A_4 + A_5 - A_6 - A_7 + A_8)\otimes \mathbb{I}=B_2\otimes \mathbb{I}\nonumber 
\end{align}
then 
\begin{equation}
		\mathcal{B}_4= 4\sum_{y=1}^{4} B_y \otimes B_y
\end{equation}
 
 It is possible to find a choice of observables and states so that each of the $\langle B_y \otimes B_y\rangle$ is $1$. In such a case, $\mathcal{B}_4=16$ providing the desired optimal probability $p^{opt}_{Q}=(1+1/2)/2$. A choice  observables and the state are the following. 

\begin{align}
\nonumber
	 B_1= \sigma_x \otimes \sigma_x, B_2= \sigma_x\otimes\sigma_y, B_3= \sigma_x\otimes\sigma_z  \text{and} \ \ B_4 = \sigma_y\otimes \mathbb{I}  
\end{align}
 and
\begin{align}
	&A_1 = \frac{1}{2}( \sigma_x \otimes \sigma_x + \sigma_x\otimes\sigma_y + \sigma_x\otimes\sigma_z + \sigma_y\otimes \mathbb{I} ) \quad 
	A_2 = \frac{1}{2}( \sigma_x \otimes \sigma_x + \sigma_x\otimes\sigma_y + \sigma_x\otimes\sigma_z - \sigma_y\otimes \mathbb{I }) \nonumber \\
	&A_3 = \frac{1}{2}( \sigma_x \otimes \sigma_x + \sigma_x\otimes\sigma_y - \sigma_x\otimes\sigma_z + \sigma_y\otimes \mathbb{I }) \quad 
	A_4 = \frac{1}{2}( \sigma_x \otimes \sigma_x - \sigma_x\otimes\sigma_y + \sigma_x\otimes\sigma_z + \sigma_y\otimes \mathbb{I }) \nonumber \\
	&A_5 = \frac{1}{2}( -\sigma_x \otimes \sigma_x + \sigma_x\otimes\sigma_y + \sigma_x\otimes\sigma_z + \sigma_y\otimes \mathbb{I} ) \quad 
	A_6 = \frac{1}{2}( \sigma_x \otimes \sigma_x + \sigma_x\otimes\sigma_y - \sigma_x\otimes\sigma_z - \sigma_y\otimes \mathbb{I }) \nonumber \\
	&A_7 = \frac{1}{2}( \sigma_x \otimes \sigma_x - \sigma_x\otimes\sigma_y + \sigma_x\otimes\sigma_z - \sigma_y\otimes \mathbb{I} ) \quad 
	A_8 = \frac{1}{2}( \sigma_x \otimes \sigma_x - \sigma_x\otimes\sigma_y - \sigma_x\otimes\sigma_z + \sigma_y\otimes \mathbb{I} ) 
\end{align}
For an entangled state $\ket{\psi}_{AB} = \dfrac{1}{2} (\ket{0000} + \ket{0101} +\ket{1010} + \ket{1111})$, the requirement of $\sum_{y=1}^{4} B_y \otimes B_y =4$ can be achieved.  

\section{}

We provide the  general construction of the observables for which the $2^{n-1}\times n$ elegant Bell's inequality is optimized. Similar construction can be found in \cite{chali}. For $n-$ bit POM task, Bob requires $n$ number of observables denoted as $B_{n,y}$ where $y\in \{1,2,....,n\}$. We already know that for 2-bit POM task $B_{2,1} = \sigma_x, B_{2,2} = \sigma_y $ and for 3-bit case $ B_{3,1} = \sigma_x, B_{3,2}=\sigma_y ,B_{3,3} = \sigma_z $. In Appendix B, we provided the Bob's observables for 4-bit POM task are $ B_{4,1} = \sigma_x \otimes \sigma_x , B_{4,2} = \sigma_x \otimes \sigma_y ,B_{4,3} = \sigma_x\otimes \sigma_z ,B_{4,4} =  \sigma_y\otimes\mathbb{I}$. Note that, $B_{n,y}$ are mutually anti-commuting. 

Using $n=3$ case, we can recursively define the observables as follows; for even  $n$, the observables are  
$B_{n,y} = \sigma_x \otimes B_{n-1,y} \;\text{for} \; y \in \{1, . . ., n-1\}, \quad B_{n,n} = \sigma_y \otimes I $ and for odd $n$, we have
$B_{n,y} = \sigma_x \otimes B_{n-2,y} \;\text{for} \; y \in \{1, . . ., n-2\}, \quad B_{n,n-1} = \sigma_y \otimes I \; \text{and} \;  B_{n,n} = \sigma_z \otimes I $.
Let Alice's observables $A_{n,i}$ can be suitably combined so that the following condition is satisfied
\begin{equation}
	\sum_{i=1}^{2^{n-1}} (-1)^{x^i_{y}}A_{n,i}\otimes\mathbb{I}=\dfrac{2^{n-1}}{\sqrt{n}}  B_{n,{y}}\otimes\mathbb{I}
\end{equation}

In such a choice of observables, the elegant Bell expression Eq.(\ref{gisinn}) can be written as  $ \mathcal{B}_{n} =  \dfrac{2^{n-1}}{\sqrt{n}}  \sum_{y=1}^{n} \langle B_{n,y} \otimes B_{n,y}\rangle$ which provides  $ \mathcal{B}_{n}^{opt}=\dfrac{2^{n-1}}{\sqrt{n}} n=2^{n-1}\sqrt{n}$. Thus, from Eq.(\ref{gisinn})  the optimal success probability for $n$-bit POM task can be written as
\begin{align}
	p_{Q}^{opt} =  \dfrac{1}{2} + \dfrac{2^{n-1}\sqrt{n}}{2^n n}  = \dfrac{1}{2} \Big( 1+ \dfrac{1}{\sqrt{n}} \Big) 
\end{align}

The entangle state of the dimension  $ 2^{\lfloor n/2\rfloor} $ provides the optimal value of the elegant Bell expression is given by

\[\ket{\phi}_{AB} = \dfrac{1}{\sqrt{2^{\lfloor n/2\rfloor}}} \sum\limits_{k=1}^{2^{\lfloor n/2\rfloor}}  \ket{k}_A \ket{k}_B \].

\end{widetext}
\end{document}